\documentclass[final,3p,times]{elsarticle}

\usepackage{amssymb}
\usepackage{amsmath}

\journal{Journal of Subatomic Particles and Cosmology}

\begin{document}

\begin{frontmatter}



\title{How strange: Phase diagrams with 3 critical points}


\author[aaa]{Mateus Reinke Pelicer}
\author[aaa]{Nikolas Cruz Camacho}
\author[ccc]{Rajesh Kumar}
\author[bbb]{Veronica Dexheimer}
\author[aaa]{Jacquelyn Noronha-Hostler}

\affiliation[aaa]{organization={The Grainger College of Engineering, Illinois Center for Advanced Studies of the Universe, Department of Physics},
             addressline={University of Illinois Urbana-Champaign},
             city={Urbana},
             postcode={61801},
             state={IL},
             country={USA}}

 \affiliation[bbb]{organization={Center for Nuclear Research, Department of Physics},
             addressline={Kent State University},
             city={Kent},
             postcode={44243},
             state={OH},
             country={USA}}

\affiliation[ccc]{organization={Department of Physics},
             addressline={MRPD Government College },
             city={Talwara},
             postcode={144216},
             state={Punjab},
             country={India}}
\begin{abstract}
We show that the Chiral Mean-Field model (CMF) can produce a phase diagram with three critical points: the nuclear liquid-gas transition, quark deconfinement, and a strangeness driven transition. The strangeness driven transition separates a mainly nucleonic phase from one dominated by hyperons and baryon resonances. We discuss the compositional change in these transitions and their possible signatures in heavy-ion collisions and neutron star mergers.
\end{abstract}




\end{frontmatter}

\section{Introduction}
\label{sec:intro}

At low temperatures and large baryon density, it is not possible to directly calculate the dense matter equation of state (EOS) from quantum chromodynamics (QCD) due to the Fermion sign problem~\cite{Troyer:2004ge}. 
Instead one relies on effective models that build in some symmetries and properties of QCD, but that remain solvable in this regime and can reproduce known theoretical and experimental constraints \cite{MUSES:2023hyz}.  
For instance, within QCD, we require three conserved charges: baryon number $B$, strangeness $S$, and electric charge $Q$, which leads to thermodynamic systems  that have three chemical potentials $\{\mu_B,\mu_S,\mu_Q\}$. 

One such model that can take this into account is the chiral mean field model (CMF) \cite{Dexheimer:2009hi} that incorporates chiral symmetry and the full baryon octet, decuplet, and quarks. 
Recently, the MUSES collaboration developed an open-source version known as CMF++ \cite{Cruz-Camacho:2024odu} at vanishing temperatures ($T=0$), which is currently being extended to finite $T$ as well. 
Within CMF++, it is possible to calculate the EOS in a 4-dimensional (4D) phase space across $\{T,\mu_B,\mu_S,\mu_Q\}$ that allows us to study different phases of matter and properties of neutron stars \cite{ReinkePelicer:2025vuh,Cruz-Camacho:2026pdg}.

Here we focus on a strangeness-dominated hadronic phase, rich in strange baryons and baryon resonances. Similar hyperon-driven first-order transitions have been found previously in effective models. We show that within CMF++ such a transition can coexist with both the nuclear liquid-gas and deconfinement transitions, leading to a phase diagram containing three distinct critical points.

\section{Strangeness dominated phase}
\label{sec:strange}

The strangeness dominated \cite{Cruz-Camacho:2024odu} phase  occurs for specific couplings within CMF when the full SU(3) baryon octet and decuplet are included.  At $T=0$, three possible states of matter can occur: a light hadronic phase (primarily composed of neutrons, protons, and sometimes $\Lambda$ baryons i.e. $np\Lambda$ matter), a quark phase, and the strangeness dominated phase ($\Xi$'s, $\Sigma$'s, and their resonances).

First-order transitions to hyperon-rich matter have been reported previously. In relativistic mean-field calculations, Schaffner-Bielich and Gal found a transition into $\Sigma$ and $\Xi$-dominated matter, while subsequent studies identified strangeness driven first-order transitions and finite-temperature critical behavior in $n\Lambda$ and $np\Lambda$ matter \cite{SchaffnerBielich:2000wj,Gulminelli:2012iq,Gulminelli:2013qr}. Full-octet calculations further showed that several hyperon-driven instabilities may occur, although their existence is strongly interaction dependent \cite{Raduta:2014lja,Torres:2016ydl}. CMF++ produces this transition alongside the liquid-gas and deconfinement transitions, giving three distinct critical points within a unified octet, decuplet and quark description.

\begin{figure}[t]
\centering
\includegraphics[width=0.45\linewidth]{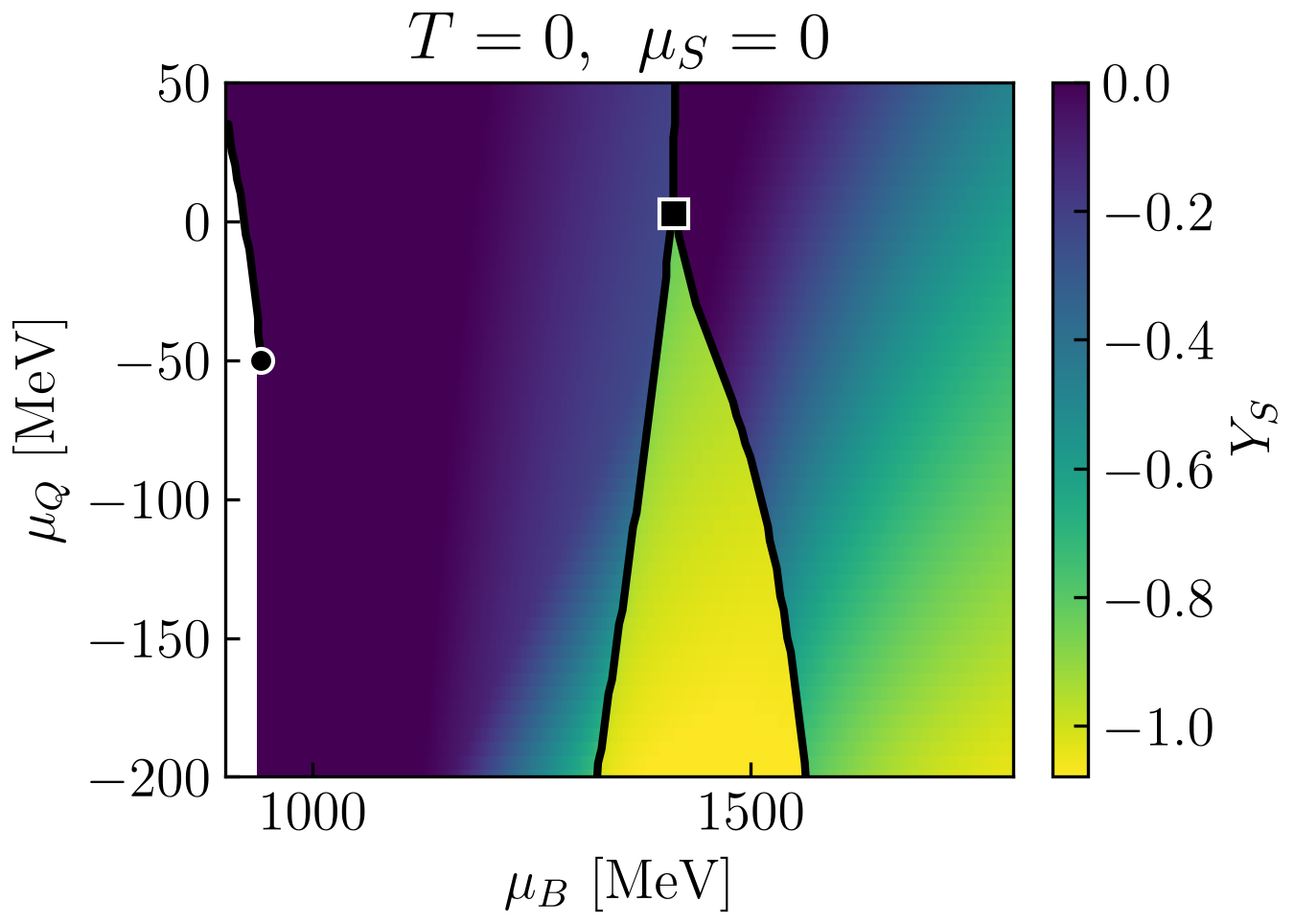}
\includegraphics[width=0.45\linewidth]{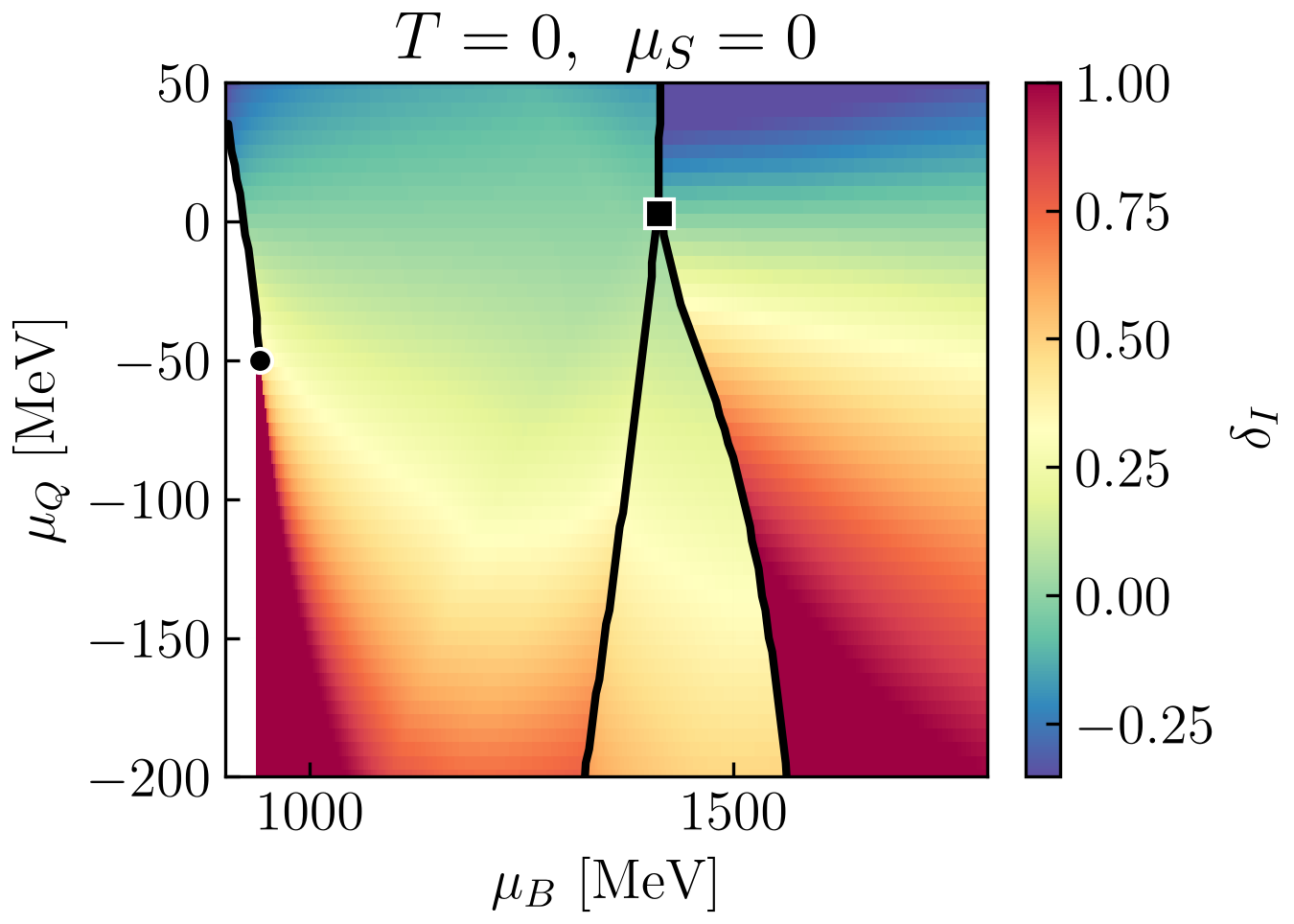}
\caption{CMF phase structure at $T=0$ for the C3 coupling. Left: strangeness fraction $Y_S$. Right: isospin asymmetry $\delta_I$. The intermediate-density strangeness-dominated phase is characterized by large strangeness and a strong reduction of the net isospin asymmetry.  }\label{fig:strange}
\end{figure}

The transition is accompanied by a pronounced rearrangement of the hadronic composition. A useful quantity for characterizing this change is the isospin asymmetry. From the Gell-Mann-Nishijima relation\cite{Yang:2025wop,Danhoni:2025qpn}
\begin{equation}
\delta_I=1-2Y_Q+Y_S
=-2\frac{n_{I_3}}{n_B},
\end{equation}
where the net electric-charge fraction is $Y_Q=n_Q/n_B$ and the strangeness fraction is $Y_S=n_S/n_B$, with strange quarks carrying $S=-1$, and  $n_{I_3} = \sum_i I_{3,i} n_i$. Thus, $\delta_I=0$ corresponds to vanishing net isospin density. For purely nucleonic matter this gives the familiar condition $Y_Q=1/2$, corresponding to equal neutron and proton abundances. Hyperonic matter, however, allows very different compositions with $\delta_I=0$. For example, equal abundances of $\Xi^-$ and $\Xi^0$ give $Y_Q=-1/2$ and $Y_S=-2$, and therefore $\delta_I=0$.

In Fig.\ \ref{fig:strange}, the light hadronic phase occurs at low $\mu_B$ across the full $\mu_Q$ range shown. The quark phase always appears at large $\mu_B$.  The strangeness dominated phase appears at intermediate $\mu_B$ and more negative $\mu_Q$. The strangeness  dominated phase is suppressed  with increasing $\mu_Q$, with a triple point between the three phases at $\mu_Q\approx 0$. On the other hand, as $\mu_Q$ decreases, the strangeness dominated phase appears at lower $\mu_B$, while the transition to quark matter moves to larger $\mu_B$. At $T=0$, this phase has $Y_S \approx -1$ and $\delta_I \lesssim 0.25$, with both quantities showing a weaker $\mu_Q$ dependence than in the light hadronic phase.

Along the $T=0$, $\mu_S=0$ slice shown in Fig.~\ref{fig:strange}, for example, $\mu_{\Xi^0}=\mu_B$ and $\mu_{\Xi^-}=\mu_B-\mu_Q$. Negative $\mu_Q$ therefore favors negatively charged baryons, including $\Xi^-$ and other strange states, shifting the relative thermodynamic preference toward the strangeness-dominated branch, while positive $\mu_Q$ suppresses it. Whether this branch forms a stable phase ultimately depends on the particle interactions.
Once strange baryons become abundant, however, the combined contributions of the different strange multiplets to $Y_Q$ and $Y_S$ can strongly reduce the net isospin density. The reduction of $\delta_I$ is therefore naturally connected to the same compositional rearrangement that produces the large strangeness fraction.

Within CMF, it is possible to implement a variety of different Ansaetze for the couplings (see \cite{Legred:2026cre} for an initial, broad exploration along those lines). Depending on the couplings, the strangeness dominated regime may be a stable phase of matter, or it could instead become a metastable phase within the deconfinement phase transition. Furthermore, the choice of coupling and corresponding parameters can strongly influence its location and predominance. Understanding this parameter dependence is therefore essential for assessing the robustness and phenomenological relevance of the strangeness driven transition.

\section{Turning on temperature}
\label{sec:finiteT}

The presence of a first-order phase transition at $T=0$ into the strangeness dominated phase naturally raises the question of whether there is an associated first-order line and critical point at finite $T$. 
Using our newly developed finite-$T$ extension of CMF++, we map this phase structure. Stable, metastable, and unstable solutions are identified using positivity of the Hessian matrix discussed in Appendix F of Ref.~\cite{Cruz-Camacho:2024odu}.

\begin{figure}[t]
\centering
\includegraphics[width=0.45\linewidth]{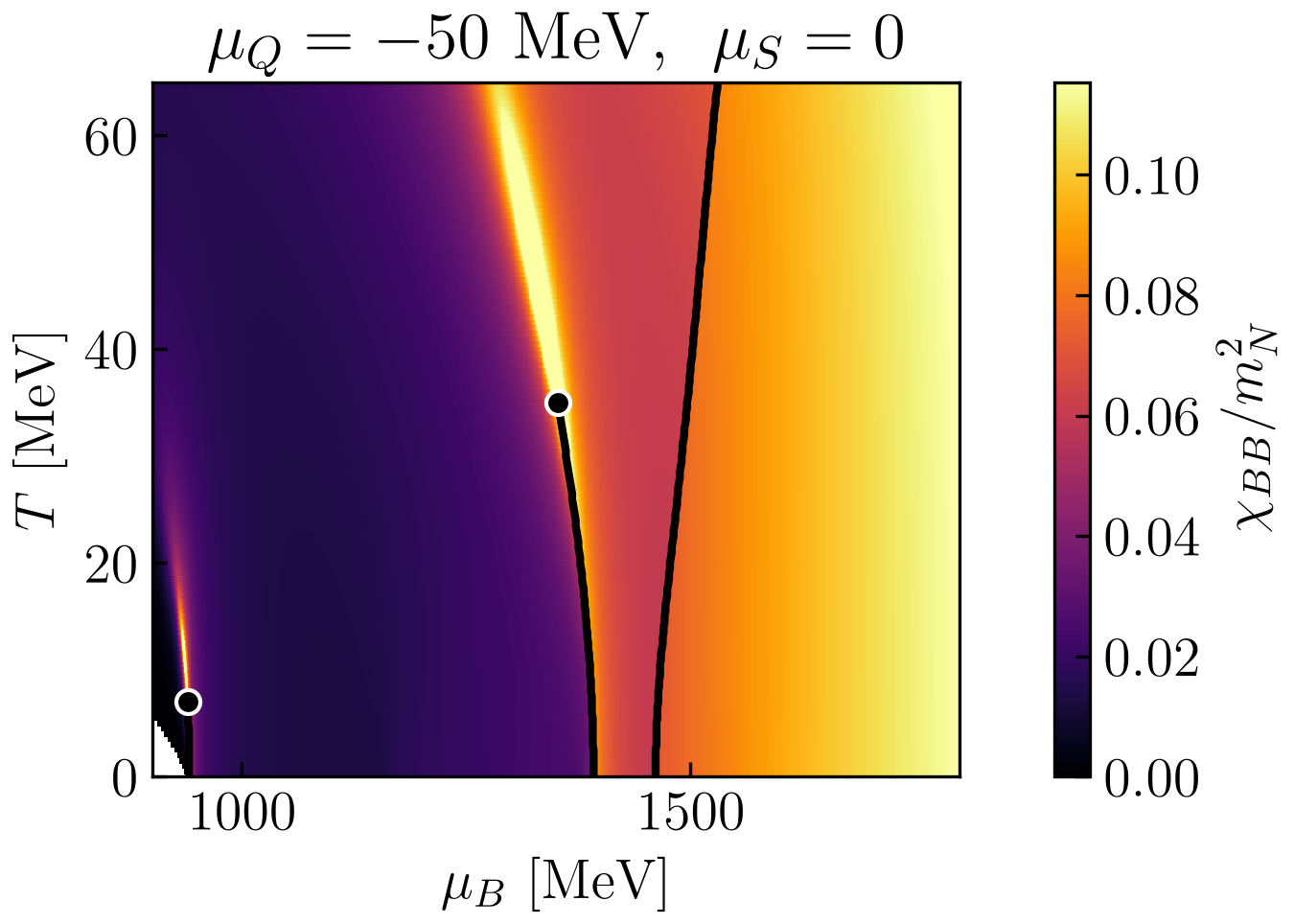}
\includegraphics[width=0.45\linewidth]{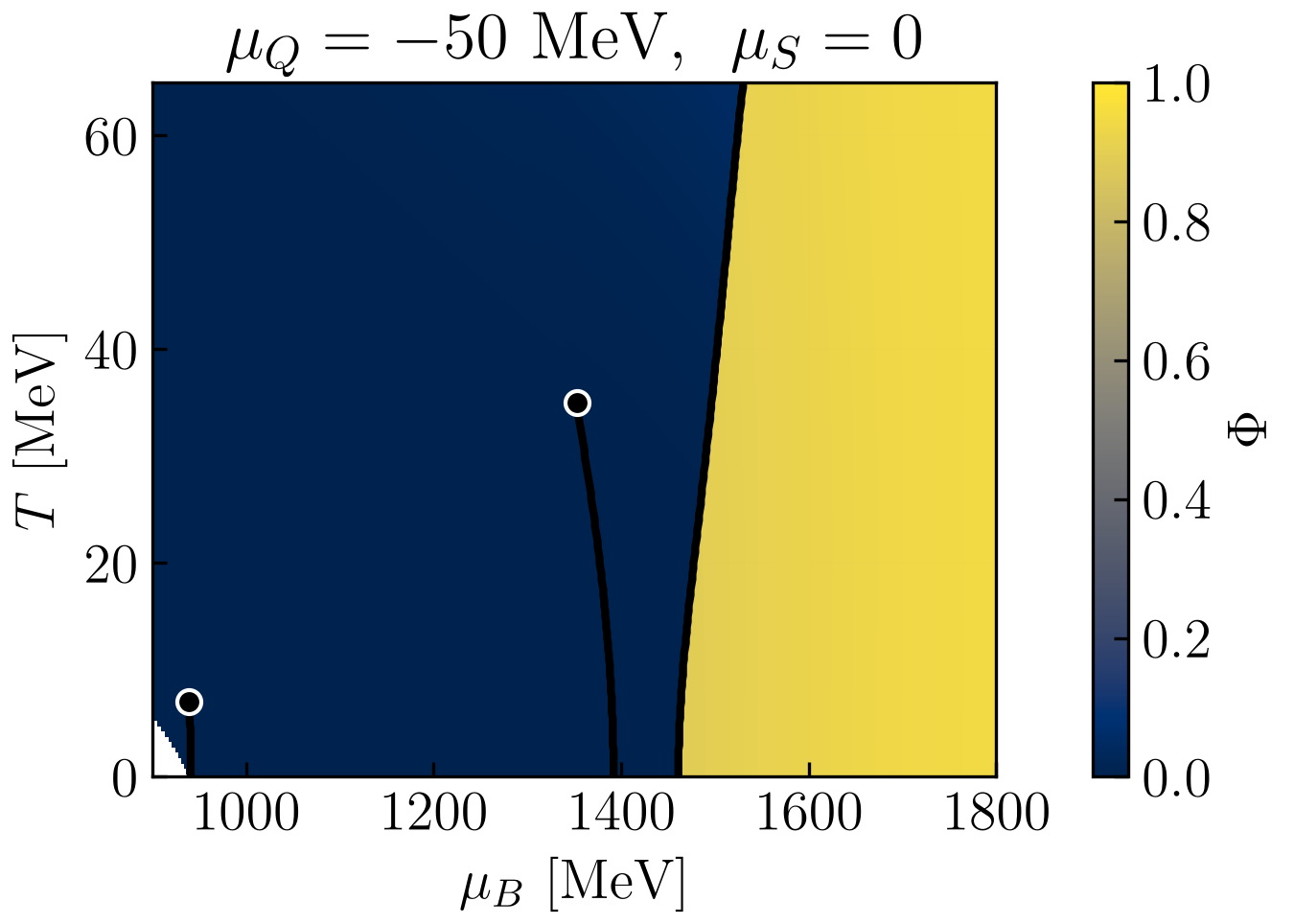}
\caption{Phase diagram of the C3 coupling EOS from CMF at $\mu_S=0$ and $\mu_Q=-50$ MeV. Left panel shows the normalized susceptibility $\chi_{BB}/m_N^2$ in color, and right panel shows the deconfinement order parameter $\Phi$.  }\label{fig:phasediagram}
\end{figure}

In Fig.\ \ref{fig:phasediagram}, we show that the C3 coupling has the three different phase transitions with corresponding critical points at finite $T$: the nuclear liquid-gas, the strangeness dominated, and the deconfinement phase transitions. The existence of the strangeness dominated critical point does not affect the nature of the liquid-gas phase transition at all. The strangeness-dominated first-order line terminates at a critical temperature of approximately $T_c\approx40$ MeV.

This temperature scale is relevant for low-energy heavy-ion collisions and hot astrophysical environments. Whether realistic trajectories present the transition depends on the appropriate charge and strangeness constraints and requires a dedicated analysis.

The normalized baryon susceptibility $\chi_{BB}/m_N^2$, shown in the left panel of Fig.~\ref{fig:phasediagram}, illustrates how the discontinuity turns into a finite broad peak associated to a continuous crossover. The right panel shows the deconfinement order parameter $\Phi$, which is only discontinuous at the hadron-quark transition, but remains $\Phi \approx 0$ in the baryonic phases.

The interplay between the strangeness dominated phase transition and deconfinement is more subtle. The deconfinement first-order transition bends towards lower $\mu_B$ at low $T$, which is opposite to the standard deconfinement line.  Further tests and explorations are warranted to determine the interplay between these regimes.

\section{Conclusions and Outlook}
\label{sec:con}

In this talk, we presented preliminary results for the strangeness-dominated first-order transition previously identified in CMF at $T=0$ and showed that it extends to finite temperature, where it terminates at a distinct critical point. Together with the nuclear liquid-gas and deconfinement transitions, this produces a phase diagram containing three critical points.

The strangeness transition has a critical point at $T_c\approx 40$ MeV and $\mu_{B_c} \approx 1345$ MeV. Its interplay with the quark phase is non-trivial:  it shifts the deconfinement first-order line to higher $\mu_B$ as temperature increases. This warrants further investigation with a wider set of CMF parameterizations. 
Future work could explore the possibility that the strangeness dominated regime could appear in low-energy heavy-ion collisions.

Ongoing work \cite{Storbacka:2026ehf} is exploring the 4D CMF++ EOS in numerical relativity simulations, and finds that it is possible to deviate from electroweak equilibrium in terms of $\mu_Q,\mu_S$ during the inspiral of binary neutron star mergers. This opens a pathway to studying how these perturbations could potentially lead to matter being perturbed into (or out of) the strangeness dominated phase within binary neutron star merger simulations. 

\emph{Acknowledgments} The authors acknowledge support from the US-DOE Nuclear Science Grant No. DE-SC0023861. 
This work was also supported in part by the National Science Foundation (NSF) within the framework of the MUSES collaboration, under grant number OAC-2103680 and from the Illinois Campus Cluster, a computing resource that is operated by the Illinois Campus Cluster Program (ICCP) in conjunction with the National Center for Supercomputing Applications (NCSA), and which is supported by funds from the University of Illinois at Urbana-Champaign.

\bibliographystyle{elsarticle-num}
\bibliography{inspire}

@article{MUSES:2023hyz,
    author = "Kumar, Rajesh and others",
    collaboration = "MUSES",
    title = "{Theoretical and experimental constraints for the equation of state of dense and hot matter}",
    eprint = "2303.17021",
    archivePrefix = "arXiv",
    primaryClass = "nucl-th",
    doi = "10.1007/s41114-024-00049-6",
    journal = "Living Rev. Rel.",
    volume = "27",
    number = "1",
    pages = "3",
    year = "2024"
}

@article{Troyer:2004ge,
    author = "Troyer, Matthias and Wiese, Uwe-Jens",
    title = "{Computational complexity and fundamental limitations to fermionic quantum Monte Carlo simulations}",
    eprint = "cond-mat/0408370",
    archivePrefix = "arXiv",
    doi = "10.1103/PhysRevLett.94.170201",
    journal = "Phys. Rev. Lett.",
    volume = "94",
    pages = "170201",
    year = "2005"
}

@article{Dexheimer:2009hi,
    author = "Dexheimer, V. A. and Schramm, S.",
    title = "{A Novel Approach to Model Hybrid Stars}",
    eprint = "0901.1748",
    archivePrefix = "arXiv",
    primaryClass = "astro-ph.SR",
    doi = "10.1103/PhysRevC.81.045201",
    journal = "Phys. Rev. C",
    volume = "81",
    pages = "045201",
    year = "2010"
}

@article{Cruz-Camacho:2024odu,
    author = "Cruz-Camacho, Nikolas and Kumar, Rajesh and Reinke Pelicer, Mateus and Peterson, Jeff and Manning, T. Andrew and Haas, Roland and Dexheimer, Veronica and Noronha-Hostler, Jaquelyn",
    collaboration = "MUSES",
    title = "{Phase stability in the three-dimensional open-source code for the chiral mean-field model}",
    eprint = "2409.06837",
    archivePrefix = "arXiv",
    primaryClass = "nucl-th",
    doi = "10.1103/PhysRevD.111.094030",
    journal = "Phys. Rev. D",
    volume = "111",
    number = "9",
    pages = "094030",
    year = "2025"
}

@article{ReinkePelicer:2025vuh,
    author = "Reinke Pelicer, Mateus and others",
    title = "{Building neutron stars with the MUSES calculation engine}",
    eprint = "2502.07902",
    archivePrefix = "arXiv",
    primaryClass = "nucl-th",
    doi = "10.1103/PhysRevD.111.103037",
    journal = "Phys. Rev. D",
    volume = "111",
    number = "10",
    pages = "103037",
    year = "2025"
}

@article{Cruz-Camacho:2026pdg,
    author = "Cruz-Camacho, Nikolas and Conde-Ocazionez, Carlos and Dexheimer, Veronica and Noronha-Hostler, Jacquelyn and Yunes, Nicol{\'a}s",
    title = "{Sensitivity of neutron star observables to microscopic nuclear parameters of realistic equations of state}",
    eprint = "2603.16019",
    archivePrefix = "arXiv",
    primaryClass = "nucl-th",
    month = "3",
    year = "2026"
}

@article{SchaffnerBielich:2000wj,
    author = "Schaffner-Bielich, Jurgen and Gal, Avraham",
    title = "{Properties of strange hadronic matter in bulk and in finite systems}",
    eprint = "nucl-th/0005060",
    archivePrefix = "arXiv",
    doi = "10.1103/PhysRevC.62.034311",
    journal = "Phys. Rev. C",
    volume = "62",
    pages = "034311",
    year = "2000"
}

@article{Gulminelli:2012iq,
    author = "Gulminelli, F. and Raduta, Ad. R. and Oertel, M.",
    title = "{Phase transition towards strange matter}",
    eprint = "1206.4924",
    archivePrefix = "arXiv",
    primaryClass = "nucl-th",
    doi = "10.1103/PhysRevC.86.025805",
    journal = "Phys. Rev. C",
    volume = "86",
    pages = "025805",
    year = "2012"
}

@article{Yang:2025wop,
    author = "Yang, Yumu and Camacho, Nikolas Cruz and Hippert, Mauricio and Noronha-Hostler, Jacquelyn",
    title = "{Symmetry-energy expansion with strange dense matter}",
    eprint = "2504.18764",
    archivePrefix = "arXiv",
    primaryClass = "nucl-th",
    doi = "10.1103/trk9-8gph",
    journal = "Phys. Rev. C",
    volume = "113",
    number = "4",
    pages = "045805",
    year = "2026"
}

@article{Danhoni:2025qpn,
    author = "Danhoni, Isabella and Yang, Yumu and Hippert, Mauricio and Noronha-Hostler, Jacquelyn",
    title = "{Symmetry energy of 2+1 -flavor dense quark matter from perturbative QCD}",
    eprint = "2510.23984",
    archivePrefix = "arXiv",
    primaryClass = "nucl-th",
    doi = "10.1103/5v39-l9lh",
    journal = "Phys. Rev. C",
    volume = "113",
    number = "4",
    pages = "045206",
    year = "2026"
}

@article{Legred:2026cre,
    author = "Legred, Isaac and Reinke Pelicer, Mateus and Dexheimer, Veronica and Noronha-Hostler, Jacquelyn and Yunes, Nicol{\'a}s",
    title = "{Neural-Accelerated Bayesian Calibration of Chiral Mean-Field Models to Nuclear Saturation and Vacuum Properties}",
    eprint = "2607.13268",
    archivePrefix = "arXiv",
    primaryClass = "nucl-th",
    month = "7",
    year = "2026"
}

@article{Storbacka:2026ehf,
    author = "Storbacka, Melvin and Wu, Jiaxi and Haber, Alexander and Most, Elias R. and Noronha-Hostler, Jacquelyn and Reinke Pelicer, Mateus and Cruz-Camacho, Nikolas and Dexheimer, Veronica",
    title = "{Strangeness Transport in Binary Neutron Star Mergers}",
    eprint = "2608.15527",
    archivePrefix = "arXiv",
    primaryClass = "astro-ph.HE",
    month = "8",
    year = "2026"
}

@article{Gulminelli:2013qr,
    author = "Gulminelli, F. and Raduta, Ad. R. and Oertel, M. and Margueron, J.",
    title = "{Strangeness-driven phase transition in (proto-)neutron star matter}",
    eprint = "1301.0390",
    archivePrefix = "arXiv",
    primaryClass = "nucl-th",
    doi = "10.1103/PhysRevC.87.055809",
    journal = "Phys. Rev. C",
    volume = "87",
    number = "5",
    pages = "055809",
    year = "2013"
}

@article{Raduta:2014lja,
    author = "Raduta, Ad. R. and Gulminelli, F. and Oertel, M.",
    title = "{Thermodynamics of baryonic matter with strangeness within non-relativistic energy density functional model}",
    eprint = "1406.0395",
    archivePrefix = "arXiv",
    primaryClass = "nucl-th",
    month = "6",
    year = "2014"
}

@article{Torres:2016ydl,
    author = "Torres, James R. and Gulminelli, Francesca and Menezes, Debora P.",
    title = "{Examination of strangeness instabilities and effects of strange meson couplings in dense strange hadronic matter and compact stars}",
    eprint = "1608.05108",
    archivePrefix = "arXiv",
    primaryClass = "nucl-th",
    doi = "10.1103/PhysRevC.95.025201",
    journal = "Phys. Rev. C",
    volume = "95",
    number = "2",
    pages = "025201",
    year = "2017"
}

\end{document}